\documentclass[journal]{IEEEtran}

\usepackage{graphicx}
\usepackage{color}
\usepackage{placeins}
\usepackage{float}
\usepackage{tabularx,colortbl}
\usepackage{amssymb}
\usepackage{amsthm}
\usepackage{cite}
\usepackage{amsmath}
\usepackage{makecell}
\usepackage{subfigure}

\usepackage{caption2}

\usepackage{algorithm}
\usepackage{algpseudocode}

\theoremstyle{plain}

\theoremstyle{plain}

\IEEEoverridecommandlockouts
\begin{document}
\title{White-Box AI Model: Next Frontier of Wireless Communications}

\author{Jiayao~Yang, Jiayi~Zhang,~\IEEEmembership{Senior Member,~IEEE}, Bokai~Xu, Jiakang Zheng, Zhilong~Liu, Ziheng Liu, Dusit~Niyato,~\IEEEmembership{Fellow,~IEEE}, M\'{e}rouane Debbah,~\IEEEmembership{Fellow,~IEEE}, Zhu Han,~\IEEEmembership{Fellow,~IEEE}, and Bo~Ai,~\IEEEmembership{Fellow,~IEEE}
\thanks{J. Yang, J. Zhang, B. Xu, J. Zheng, Z. Liu, Z. Liu and B. Ai are with the School of Electronic and Information Engineering, Beijing Jiaotong University; D. Niyato is with Nanyang Technological University;
M. Debbah is with KU 6G Research Center, Department of Computer and
Information Engineering, Khalifa University;
Z. Han is with University of Houston.}
}
\maketitle
\vspace{-1.8cm}

\begin{abstract}
White-box AI (WAI), or explainable AI (XAI) model, a novel tool to achieve the reasoning behind decisions and predictions made by the AI algorithms, makes it more understandable and transparent. It offers a new approach to address key challenges of interpretability and mathematical validation in traditional black-box models. In this paper, WAI-aided wireless communication systems are proposed and investigated thoroughly to utilize the promising capabilities. First, we introduce the fundamental principles of WAI. Then, a detailed comparison between WAI and traditional black-box model is conducted in terms of optimization objectives and architecture design, with a focus on deep neural networks (DNNs) and transformer networks. Furthermore, in contrast to the traditional black-box methods, WAI leverages theory-driven causal modeling and verifiable optimization paths, thereby demonstrating potential advantages in areas such as signal processing and resource allocation. Finally, we outline future research directions for the integration of WAI in wireless communication systems.
\end{abstract}
\begin{IEEEkeywords}
White-box AI model, wireless communication, information theory, optimization, high-dimensional statistics.
\end{IEEEkeywords}

\section{Introduction}\label{sec1}
The sixth-generation (6G) wireless communication systems aim to meet the increasing demand for ultra-high data rates, massive connectivity, and enhanced spectral and energy efficiency, thereby driving further advancements in enabled technology\cite{7,1}. As the complexity and dynamics of communication systems continue to evolve, the integration of intelligent, adaptive, and efficient optimization frameworks becomes essential to ensure network reliability and scalability. Recently, artificial intelligence (AI) has emerged as a core solution to these challenges, making significant strides in signal processing and resource allocation\cite{8}.

Deep neural networks (DNNs), residual Networks (ResNets), and Transformer have demonstrated significant potential in optimizing wireless networks\cite{10,2}. Neural networks identify complex signal transmission patterns and make optimization predictions by efficiently learning from historical data. Transformer-based large AI models are particularly recognized for their advantages in processing sequential information and parallel computing, enabling the efficient handling and scheduling for high-speed data flows\cite{10}. However, despite these advances, the dominance of black-box models in wireless communication remains. Although these models are effective in many scenarios, they often lack transparency and interpretability. This limitation has driven the development of white-box AI (WAI) models, which offer greater explainability and mathematical verifiability. 

    
\begin{figure*}[t]
	\centering
	\includegraphics[width=0.9\textwidth]{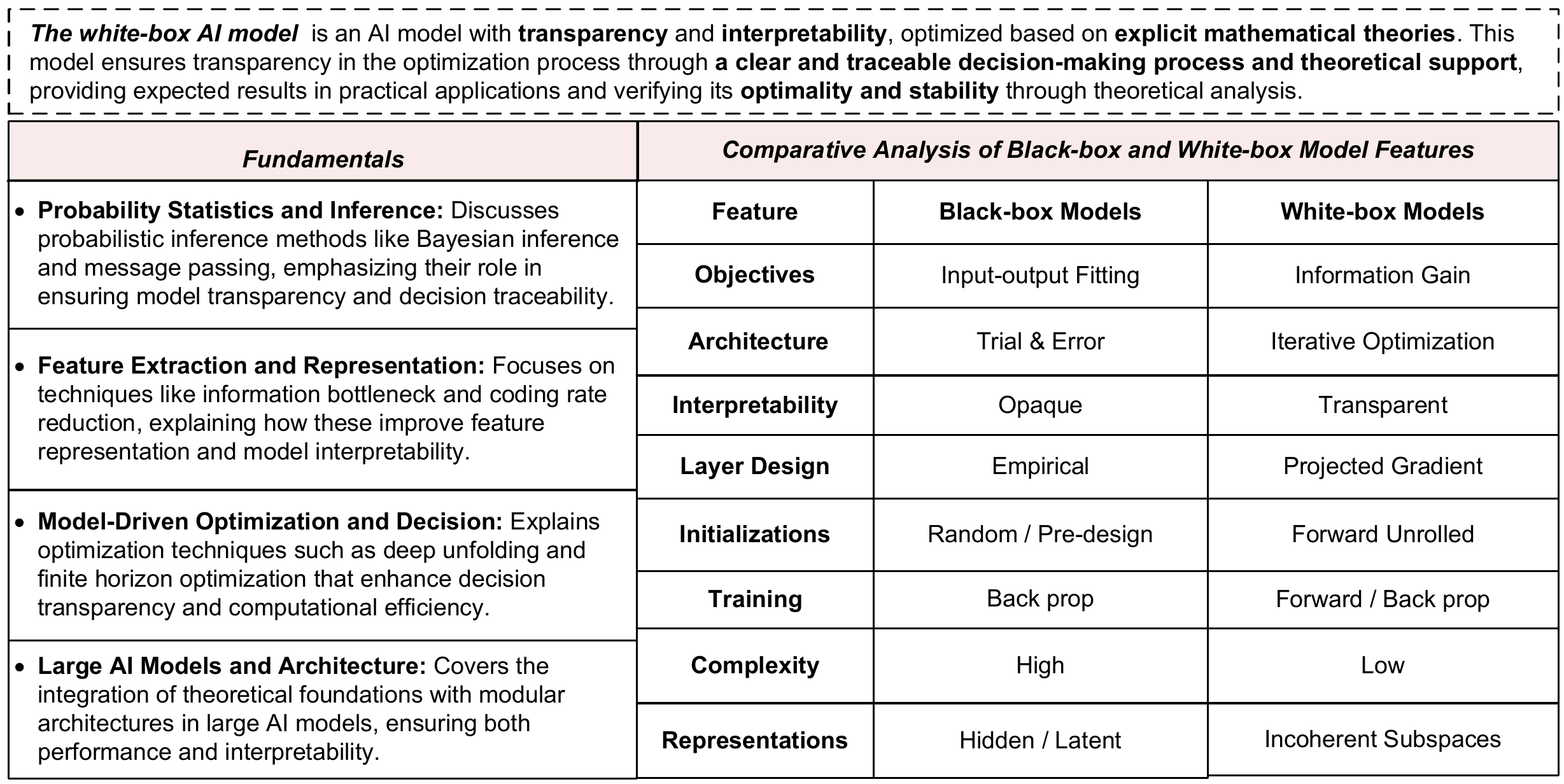}
	\caption{Overview of WAI model: definition and fundamentals, with a comparative analysis of white-box and black-box model features.} \label{1}
\end{figure*}
$\bullet$ \textbf{\emph{Traditional Black-Box Network:}}
Despite significant advancements in AI for wireless communication, the application of black-box models still faces several challenges. i) Lack of interpretability and theoretical foundation: Due to the end-to-end learning nature, the decision-making process remains opaque to users, thus making it difficult for users to understand and interpret the optimization process and its results. This opacity limits performance analysis, engineering implementation, and practical validation in wireless communication systems. ii) High computational complexity: Tasks such as large-scale signal processing and resource allocation typically involve high-dimensional non-convex optimization problems. However, these tasks depend on extensive computational resources, resulting in suboptimal performance in time-sensitive scenarios. iii) Limited generalization ability: Especially in scenarios involving dynamic channel variations, non-stationary interference, and unpredictable user behavior, the adaptability of black-box models is limited. 


$\bullet$ \textbf{\emph{White-Box Network:}} 
To overcome the limitations of black-box models, WAI models have emerged as a new modeling paradigm in wireless communication systems. The core advantage of WAI lies in its transparency and mathematical verifiability, which render it particularly suitable for wireless communication environments that require high reliability and interpretability. Traditional black-box models, such as DNN, lack explicitly interpretable optimization paths, making it difficult to provide clear decision rationales. This limitation affects the feasibility of engineering implementation and increases the uncertainty of optimization. In existing research, physics-informed machine learning (PIML) incorporates physical laws and mathematical principles to enhance model interpretability\cite{13}. However, PIML and the proposed WAI differ fundamentally in methodology and application. PIML focuses on physical modeling, integrating known physical laws to constrain neural network learning, and is widely applied in fields such as fluid dynamics and weather prediction. In contrast, WAI integrates information theory and optimization theory, establishing a mathematically interpretable optimization framework\cite{1}. This ensures transparent and theoretically verifiable optimization paths, enhancing adaptability in complex and dynamic network environments\cite{2}. In Fig. \ref{1}, we summarize the basic fundamentals  of the WAI model. Moreover, we compare the features and architectures of white-box and black-box. The major advantages of the WAI are as follows: 

$\bullet$ {\bf Explainablity of WAI: }The WAI emphasizes transparency and theoretical verifiability in optimization. Unlike conventional black-box AI, whose decision-making processes are difficult to interpret, WAI explicitly reveals its internal mechanisms and optimization logic. By integrating theoretical constraints such as information theory and mathematical optimization, it ensures clarity and interpretability of each decision, thereby making the optimization path transparent and mathematically verifiable.

$\bullet$ {\bf Reliability of WAI: }The WAI enhances decision reliability by leveraging theory-driven learning rather than purely data-driven methods. While traditional black-box models require extensive training data and are sensitive to data quality and environmental variations, WAI incorporates well-established theoretical frameworks. For example, by employing information bottleneck (IB) principles in cell-free massive multiple-input multiple-output (mMIMO) systems, WAI efficiently extracts critical information from high-dimensional, correlated channel state information (CSI), filtering out redundant interference and noise\cite{5}. This approach significantly improves model robustness and reliability in dynamic environments.

$\bullet$ {\bf Sustainability of WAI: }Sustainability is another key feature, emphasizing a balanced trade-off between computational efficiency and performance enhancement. WAI leverages adaptive optimization strategies and efficient algorithmic structures  to achieve quick adaptability across varying wireless scenarios, thus reducing computational complexity and resource consumption. This balance ensures the model operates efficiently, sustainably managing computational resources while achieving robust performance.

Based on the above characteristics of WAI, this paper aims to offer a beginner’s guide to the next-generation wireless communications aided by WAI. To this end, we first comprehensively introduce the core principles of WAI, emphasizing its transparency and interpretability in Section \ref{sec2}. Then, we explore the application of WAI in wireless communication and discuss how WAI models optimize key tasks, highlighting their advantages over traditional black-box models in Section \ref{sec3}. Subsequently, we demonstrate a practical application of WAI through case studies in Section \ref{sec4}. Last, we highlight several wireless communication-related open problems as important future directions to research WAI in Section \ref{sec5}.

    

\begin{figure*}[t]
	\centering
	\includegraphics[width=1\textwidth]{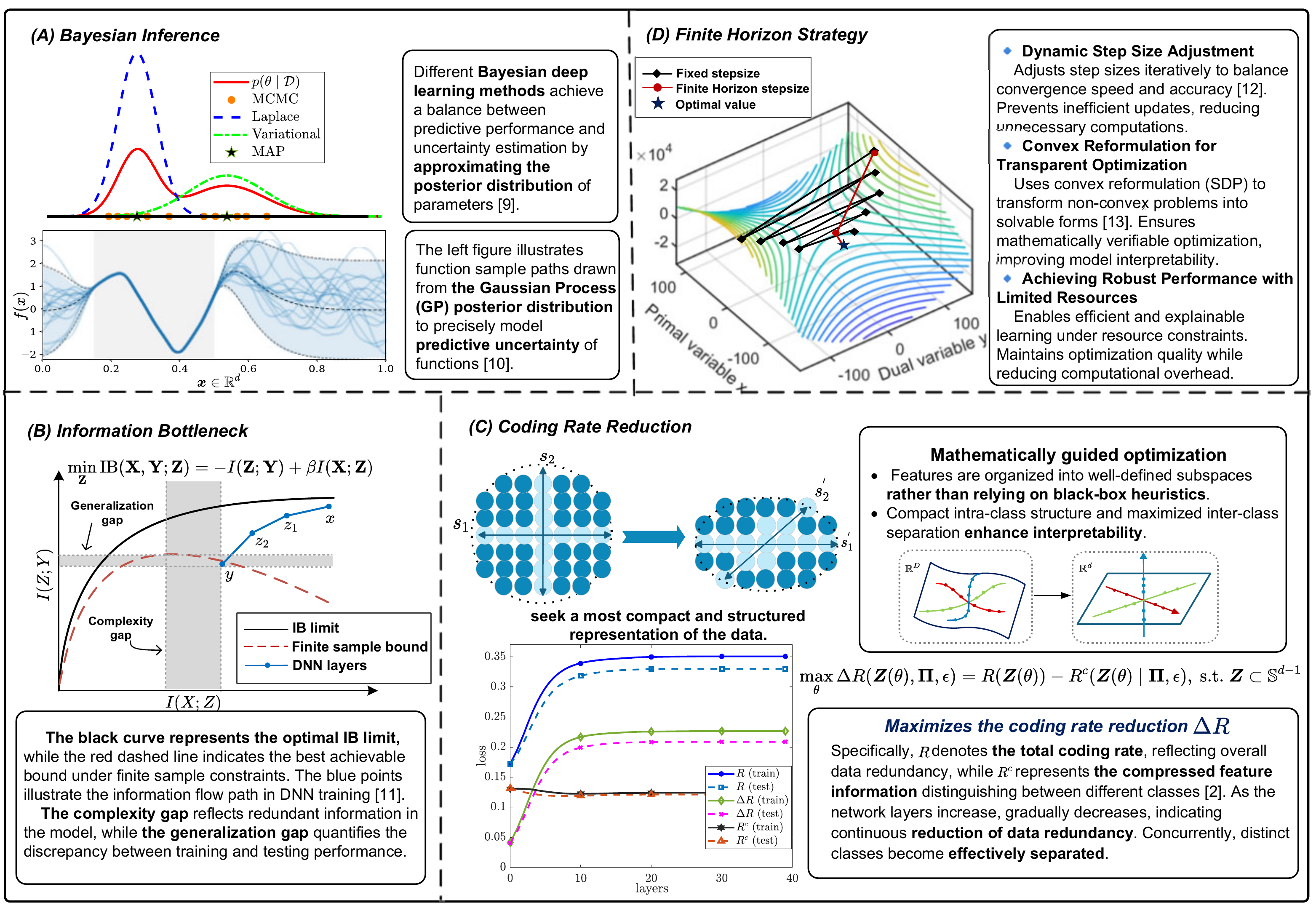}
	\caption{Fundamental theories supporting WAI model: (A) Bayesian Inference, (B) Information Bottleneck, (C) Coding Rate Reduction, and (D) Finite horizon strategy. } \label{2} 
\end{figure*}

\section{Fundamentals of Explainability in AI Models}\label{sec2}
In this section, we explore the core dimensions of explainable modeling and systematically analyze the four foundational directions that support the development of WAI.

\subsection{Probability Statistics and Inference}
This section takes a probabilistic perspective to explore the advantages of probabilistic inference mechanisms in model transparency and the traceability of the inference process. These mechanisms ensure high inference performance while significantly enhancing model interpretability.
\subsubsection{Bayesian Inference}
Bayesian inference methods are built on a rigorous probabilistic modeling framework. The inference process is structured through prior distributions, likelihood functions, and posterior updates. Unlike traditional black-box models, which rely on implicit mappings from training data and model parameters, Bayesian inference is driven by mathematical formulas\cite{15}. Each update step has a clear statistical interpretation, ensuring high transparency and traceability in the process. Fig. \ref{2} (A) reveals that various Bayesian methods, such as Markov Chain Monte Carlo (MCMC), approximate the posterior distribution to perform predictions and uncertainty estimation\cite{14}. Additionally, by sampling multiple function paths from the posterior distribution of a Gaussian Process, the model clearly demonstrates the predictive confidence in different input regions\cite{12}. This further highlights the crucial role of Bayesian inference in WAI reasoning.
\begin{figure*}[h]
	\centering
	\includegraphics[width=0.85\textwidth]{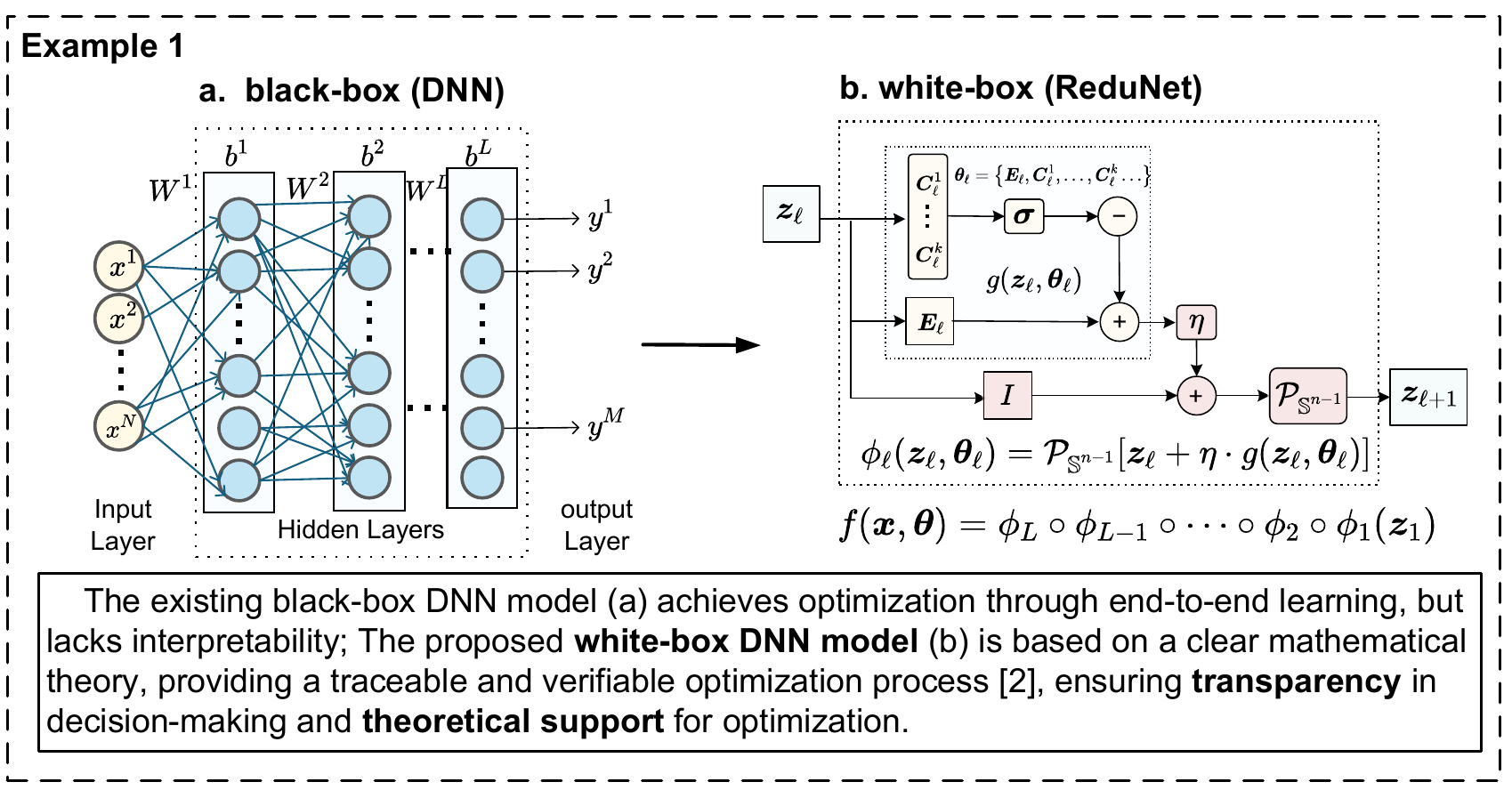}
	\caption{ReduNet: A WAI model as an alternative to traditional black-box DNN model.} \label{3} 
\end{figure*}
\subsubsection{Message Passing}
Message passing is a widely used inference mechanism in graphical models, including Bayesian networks and Markov random fields. Its fundamental idea is to approximate the joint posterior distribution by iteratively passing local information in a factor graph. A representative method is belief propagation, which updates messages between variable and factor nodes to efficiently propagate information through the graph. Compared to traditional black-box models, message passing has a well-defined computational structure and an interpretable flow of information. Both the inference path and the update rules are traceable and transparent, demonstrating high structural interpretability.

\subsection{Feature Extraction and Representation}
This section discusses how to compress input redundancy and extract key features from the perspective of information theory. It explains the model’s decision-making process and the underlying representation mechanisms.
\subsubsection{Information Bottleneck}
IB theory provides a mathematical framework to optimize the hidden representation $\mathbf{Z}$, which captures the most relevant features of the signal while suppressing redundant information\cite{9}. As shown in Fig. \ref{2} (B), the principle is to retain sufficient information about the target $\mathbf{Y}$ while removing redundancy from the input $\mathbf{X}$. $I\left(\mathbf{Z} ; \mathbf{Y}\right)$ quantifies the information provided by $\mathbf{Z}$ about $\mathbf{Y}$, and $I\left(\mathbf{X}; \mathbf{Z}\right)$ measures the retained information from $\mathbf{X}$. The parameter $\beta$ controls the trade-off between relevance and compression. This theory explicitly models the information relationship between features and the target, ensuring that the model’s representation process has a clear reasoning path and interpretability.

\subsubsection{Coding Rate Reduction}
Rate reduction provides a robust framework for learning feature representations. It focuses on two primary goals: ensuring compactness within structured feature subsets and maximizing the separation between these subsets \cite{1}. As seen in Fig. \ref{2} (C), this principle aims to maximize the difference between the total coding rate $R(\mathbf{Z})$ and the conditional coding rate $R^c(\mathbf{Z} \mid \mathbf{\Pi})$. The total coding rate $R(\mathbf{Z})$ represents the overall capacity of the learned features, while the conditional coding rate $R^c(\mathbf{Z} \mid \mathbf{\Pi})$ controls the compactness of information within the structured feature subsets. Here, $\mathbf{Z}$ denotes the learned feature representations, and $\mathbf{\Pi}$ is a set of diagonal matrices representing the soft assignment of samples to subspaces. This principle organizes similar features into compact subspaces while ensuring distinct separation among different subsets.  The resulting structured arrangement minimizes redundancy, improves feature separability, and enhances the interpretability of learned representations.

 {\bf Example 1: } As a specific implementation of the coding rate reduction principle, ReduNet serves as an interpretable example of a WAI model\cite{1}. As illustrated in Fig. \ref{3}, ReduNet implement this principle through forward optimization. Features are refined iteratively using the update rule. Specifically, each layer explicitly computes the total coding rate $R$ and class-specific coding rate $R^{c}$, updating parameters by optimizing their difference. Unlike traditional networks relying on backpropagation, ReduNet derives parameters explicitly from feature statistics. This explicit computation provides clear mathematical meaning for each layer.

\begin{figure*}[t]
	\centering
	\includegraphics[width=0.9\textwidth]{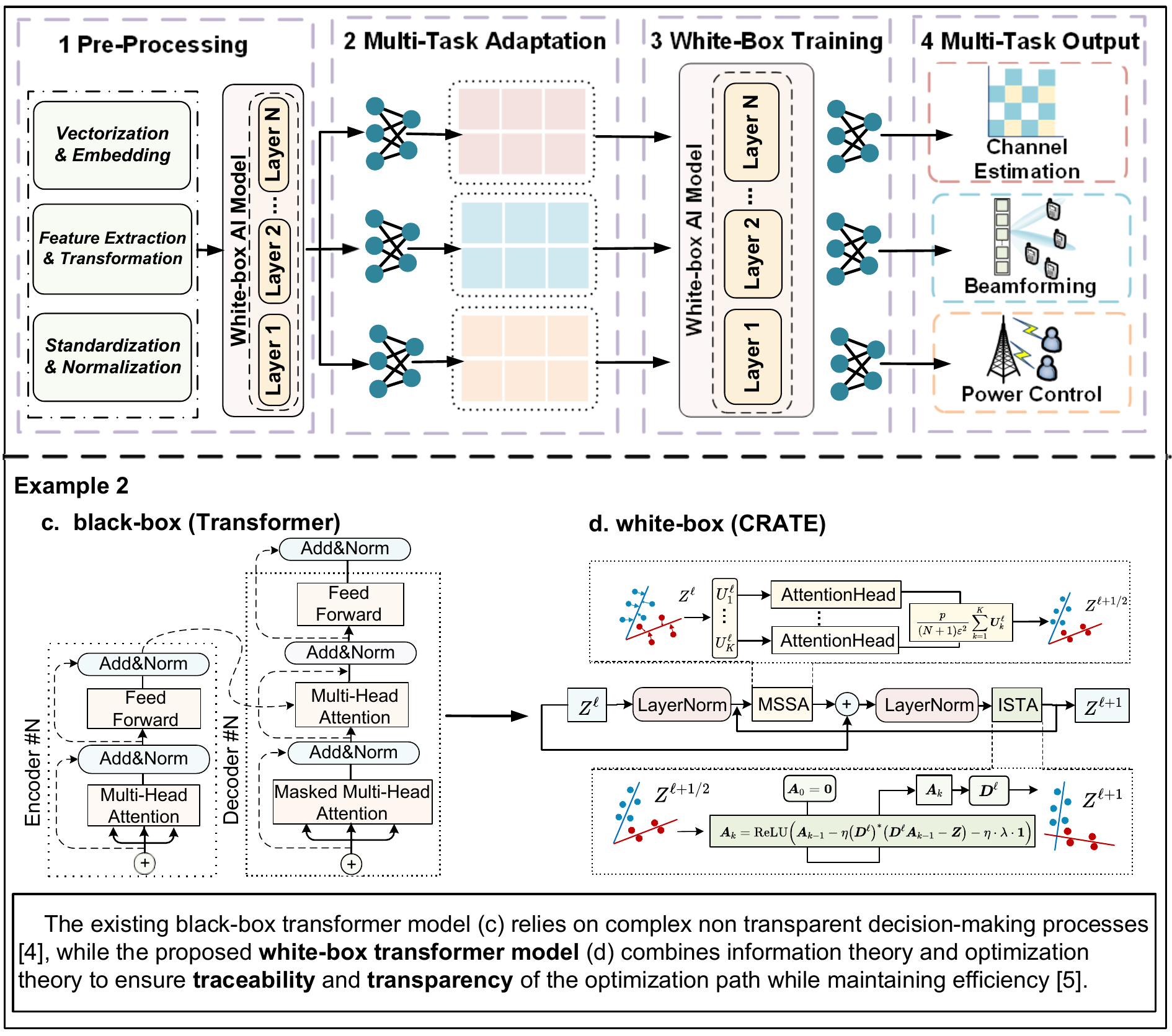}
	\caption{Architecture of the white-box wireless large AI model. The model comprises four core modules: pre-processing, multi-task adaptation, white-box training, and multi-task output. With the Transformer serving as a representative example, illustration of its white-box counterpart, CRATE.} \label{4}
\end{figure*}
\subsection{Model-Driven Optimization and Decision}
This section focuses on model-driven optimization methods. The goal is to achieve efficient and interpretable decision-making. It highlights the interpretability of optimization paths and the transparency of the computational process.
\subsubsection{Deep Unfolding}
Deep unfolding is a technique that combines deep learning with iterative optimization algorithms. It preserves the structure of traditional optimization processes, where each update rule and computational step can be traced and explained through mathematical formulas\cite{4}. This ensures transparency and traceability of the optimization process. Through deep unfolding, the model learns the optimal parameters for each iteration, improving both convergence speed and optimization efficiency. Unlike traditional methods, deep unfolding combines the advantages of model-driven optimization with the computational power of deep learning. This allows the optimization process to both explain the rationale behind each decision and leverage data-driven approaches to solve problems efficiently. This method supports WAI models by ensuring both interpretability and efficiency.
\subsubsection{Finite Horizon Optimization}
Finite horizon optimization focuses on optimizing algorithm performance within a finite iteration budget. As shown in Fig. \ref{2} (D), its core objective is to minimize the algorithm's worst-case error by dynamically adjusting parameters, such as step sizes, under strict iteration constraints\cite{3}. It adjusts step sizes iteratively to optimize the cumulative effect of matrix updates, controlling the singular value spectrum of the asymmetric update matrix and accelerating error decay. However, the asymmetry of the update process and the dynamic unpredictability of the projection operations increase complexity, making the problem inherently non-convex. 
To address this, the framework introduces hidden convexity by reformulating the problem as a semidefinite programming (SDP) problem, reconstructing it into a convex form that can be efficiently solved\cite{3}. This ensures theoretical transparency and high computational efficiency.

\begin{figure*}[t]
	\centering
	\includegraphics[width=0.9\textwidth]{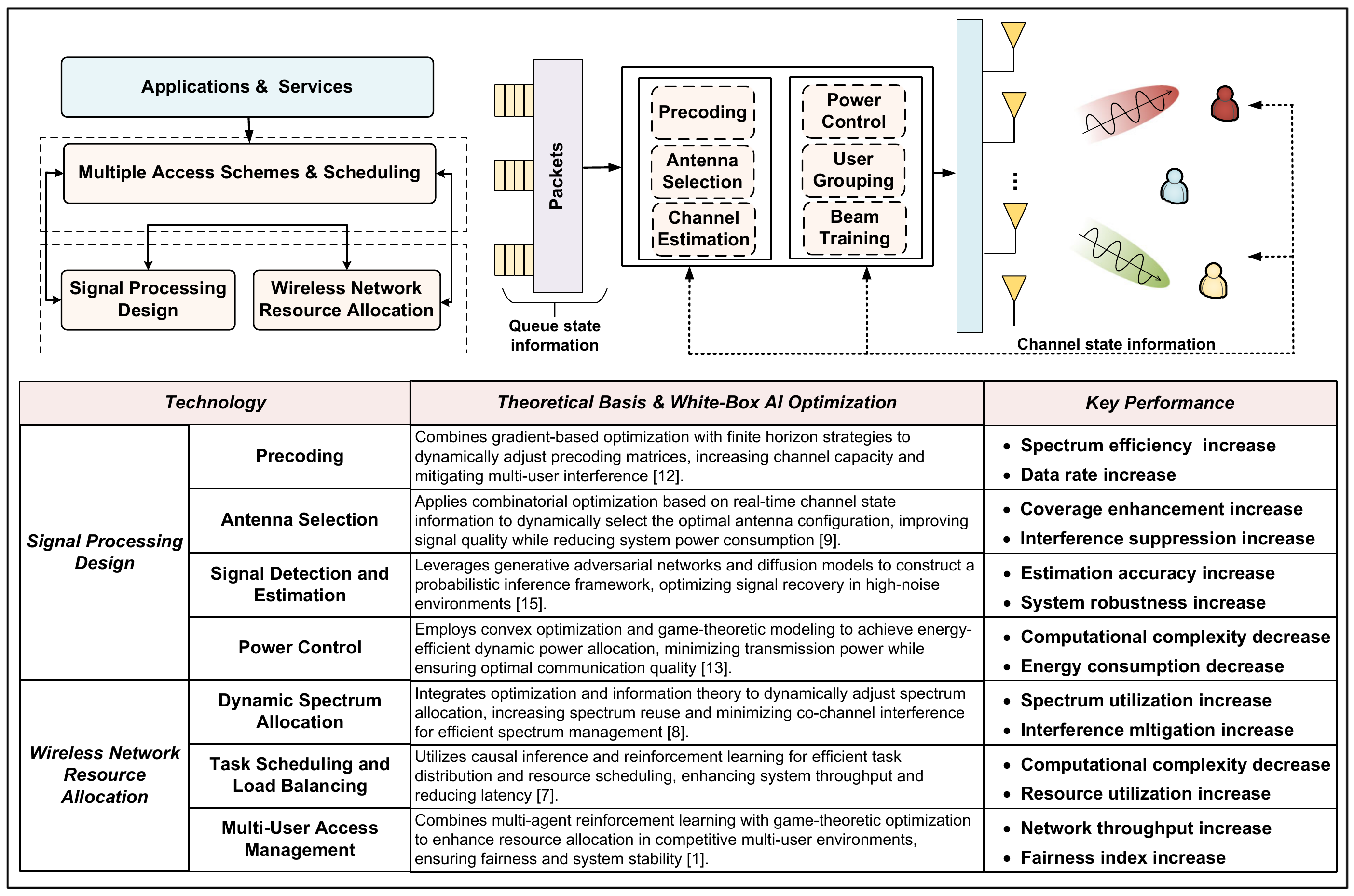}
	\caption{Applications of WAI in wireless communication: architecture and optimization techniques. Key modules include signal processing and wireless resource management. Furthermore, we provide an overview of the WAI optimization design for core technologies in wireless communication.} \label{5}   
\end{figure*}
\subsection{Large AI Model and Architecture}
The white-box wireless large AI model moves beyond traditional black-box AI by integrating a structured optimization framework grounded in mathematical principles. This design enhances transparency and interpretability while ensuring optimization follows theoretical constraints. To adapt to diverse communication tasks, the model employs a modular architecture, as shown in Fig. \ref{4}, consisting of the following four core components\cite{11}: \textit{(1) Pre-process:} This module performs input normalization, key feature extraction, noise suppression, and redundancy reduction. Each step, such as vectorization and feature transformation, follows a mathematically grounded and interpretable process. The entire pipeline is traceable and provides a stable and efficient foundation for subsequent modeling and optimization. \textit{(2) Multi-task adapter:} This module dynamically adjusts intermediate representations and parameter configurations to support various tasks, including channel estimation, beamforming, and power allocation. The task-adaptive mechanism enables effective semantic alignment and task-specific adaptation, significantly enhancing the model's generalization ability and flexibility across diverse communication scenarios. \textit{(3) White-box training:} This module integrates principles from information theory, optimization, and probabilistic modeling to construct a mathematically verifiable training framework. During training, decision steps follow clearly defined optimization paths, and parameter updates in each layer are guided by theoretical rules.  \textit{(4) Multi-task output:} This module generates optimal control strategies and system parameters based on the requirements of each communication task. It aligns task performance with model interpretability and supports a wide range of wireless communication applications.

{\bf Example 2: }It is widely recognized that Transformers serve as a fundamental building block of the large AI model\cite{11}.  We focus on the Transformer as a representative example to introduce its white-box counterpart in Fig. \ref{4}, namely CRATE\cite{2}. CRATE is designed based on the principle of maximizing coding rate reduction, enabling a structured and interpretable optimization process within Transformer architectures. Its core components include Multi-Head Subspace Self-Attention (MSSA), which captures multi-scale features and enhances information extraction through subspace decomposition, and the Iterative Shrinkage-Thresholding Algorithm (ISTA), which refines sparse features to improve compactness and discriminability\cite{2}. This structured design ensures that CRATE maintains a mathematically interpretable optimization process within the WAI framework.

\section{Application for wireless Communication}\label{sec3}
In this section, we explore the application of WAI in wireless communication, with a particular focus on the two core areas of signal processing and resource allocation.
\subsection{Signal Processing}
\subsubsection{Channel Estimation and Detection} 
WAI models employ probabilistic modeling to define the statistical relationships between signals and interference, ensuring that each optimization step is grounded in a solid theoretical foundation. Variational inference further refines the posterior distribution, enhancing the mathematical verifiability and convergence of the channel estimation process\cite{15}. By incorporating prior knowledge, WAI models adaptively estimate the channel state in dynamic environments, thus improving estimation accuracy and mitigating the over-reliance on training data typical of black-box models. In signal detection, WAI models approach the process as posterior inference, dynamically estimating the posterior probabilities of transmitted symbols, which addresses the limitations of fixed decision boundaries in traditional methods\cite{14}. Additionally, generative models, such as diffusion models, simulate the signal generation process, improving signal recovery accuracy, especially in high-interference or non-Gaussian environments\cite{6}.

\subsubsection{Precoding and Power Control} 
In precoding, the IB principle provides an efficient optimization framework\cite{5}. It optimizes signal representation using clear mathematical theory, retaining the most relevant features while reducing redundancy. This enables the system to better adapt to complex channel environments and enhances its performance. WAI models use deep unfolding for adaptive optimization\cite{4}. With gradient descent, the model adjusts the precoding matrix based on local feedback during each iteration, optimizing the relationship between signals and interference. By adjusting power allocation, WAI models improve transmission efficiency under resource constraints and enhance system robustness.

\subsection{Resource Allocation}
\begin{figure*}[h]
	\centering
\includegraphics[width=0.8\textwidth]{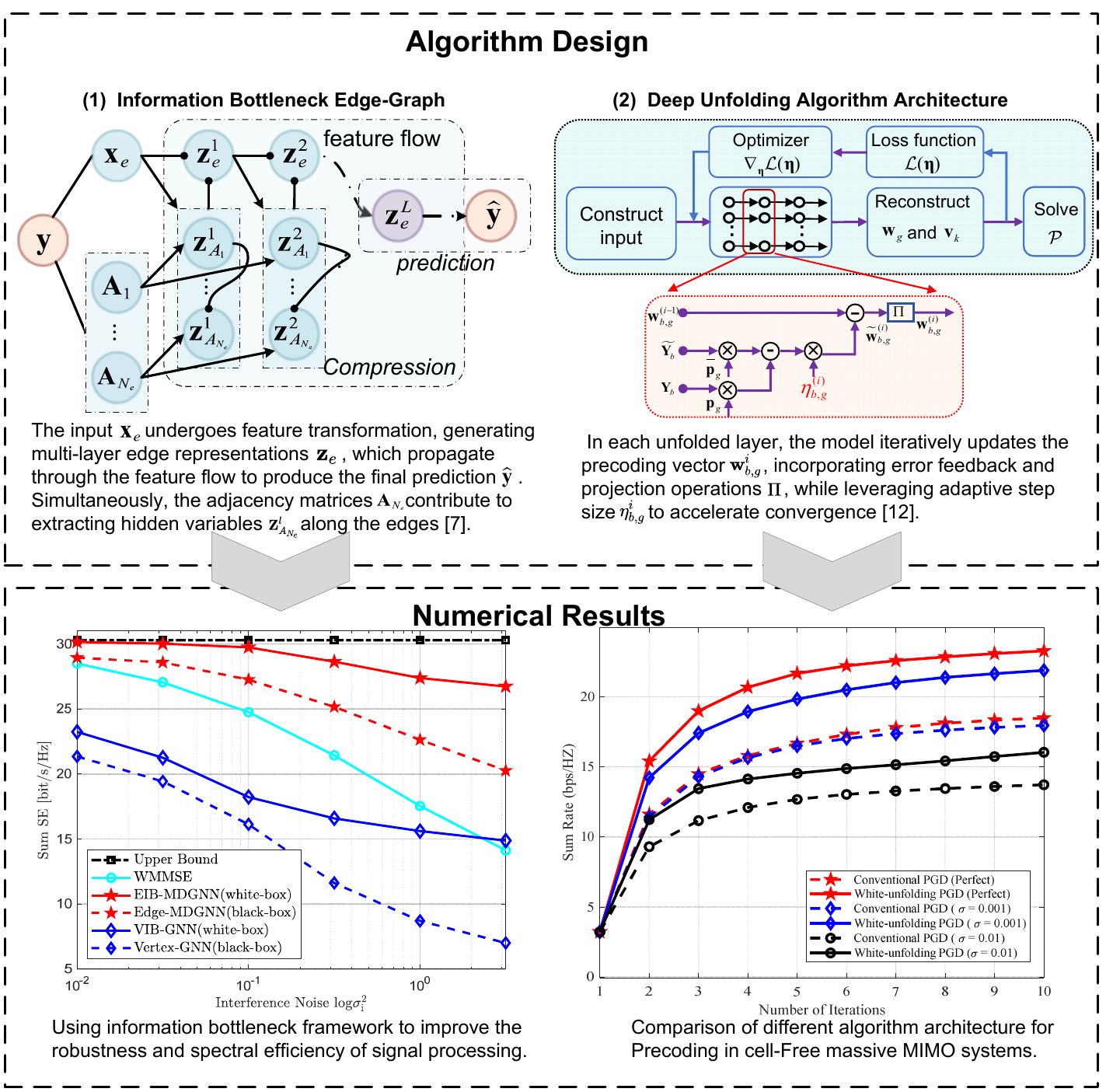}
	\caption{Algorithm design and performance evaluation of precoding optimization in WAI for cell-free mMIMO systems: (1) Information bottleneck-based optimization, (2) Deep unfolding-based optimization.} \label{6}
\end{figure*}
\subsubsection{Dynamic Resource Allocation}In dynamic resource allocation, WAI models combine finite horizon optimization and probabilistic inference to optimize the distribution of computing, storage, and network resources, improving system adaptability and resource utilization efficiency\cite{4}. Through finite horizon optimization, WAI models dynamically adjust step sizes and resource allocation strategies at each optimization step. This helps minimize worst-case error while balancing performance and resource consumption within a limited time budget. Probabilistic inference models the uncertainty in network status, device demand, and computational load, allowing the system to adjust resource allocation adaptively based on real-time information\cite{12}. 


\subsubsection{Task Scheduling and Load Balancing}Task scheduling mechanisms shift from experience-driven methods to model-based optimization. Through a causal reasoning framework, the system can deeply analyze the relationship between network load and service demands, dynamically adjusting spectrum resources and latency requirements\cite{8}. This approach effectively addresses the instability of existing load balancing algorithms, ensuring a stable balance between service quality and resource utilization, especially in ultra-dense network environments.
\subsubsection{Multi-user Access Management}By combining game theory and information theory, a new resource management framework is developed based on the white-box concept\cite{2}. It models the competitive and cooperative relationships between user devices, breaking the limitations of traditional technologies and supporting more efficient user access. Through a multi-agent collaborative distributed decision-making architecture\cite{7}, the framework ensures fairness and auditability in resource allocation, improving the rationality and reliability of wireless resource distribution.

To sum up, we present a comprehensive overview of the applications of WAI in wireless communication, covering its architectural design, key application scenarios, and core optimization techniques, as shown in Fig. \ref{5}.

\section{Case Studies}\label{sec4}
This section evaluates and demonstrates the WAI in wireless communication, with a focus on precoding optimization in cell-free MIMO systems.

\textbf{Case 1:} 
We perform the simulation analysis to evaluate the improvement in total spectrum efficiency (SE) within wireless communication systems, particularly under conditions of high interference noise. Additionally, we employ an optimization approach grounded in the IB framework to enhance system performance. The simulation adopts a cell-free massive MIMO architecture, where $M=10$ access points (APs) are deployed. Each AP is equipped with $N=4$ antennas and simultaneously serves $K=4$ single-antenna user equipments (UEs). All APs are connected to a central processing unit via fronthaul links. Signal transmission is affected by the Rayleigh fading channel model, with a communication bandwidth of 20 MHz, and each AP has a maximum transmission power limit of 1 W. The core issue addressed in this study is how to optimize precoding and power control to maximize the system's total SE in high-noise interference environments.

We propose an edge graph information bottleneck (EGIB)-based multidimensional graph neural network (EIB-MDGNN) optimization framework\cite{5}. Unlike traditional black-box GNN approaches, this method explicitly models and optimizes the information flow, ensuring that feature extraction and optimization paths have clear mathematical interpretations\cite{9}. Specifically, the framework employs hyper-edges as information interaction units, dynamically updating their hidden representations to mitigate information loss. Simultaneously, it leverages the IB principle to retain only task-relevant information while effectively suppressing redundant or irrelevant input data. This transparent optimization mechanism not only enhances the model's robustness and generalization capability but also improves the interpretability of parameter updates, providing a theoretically grounded approach for efficient signal processing in wireless communication systems.

From Fig. \ref{6}, we can observe that the proposed EIB-MDGNN method outperforms traditional baselines such as WMMSE and various GNN frameworks, showing greater robustness in handling random interference noise under practical conditions. In particular, as interference noise gradually increases, the performance gap between the proposed optimization frameworks and traditional baselines widens. This optimization method not only improves signal processing efficiency but also reduces computational resource requirements, demonstrating higher adaptability and system transparency. 

\textbf{Case 2:} 
As shown in Fig. \ref{6}, we investigate the application of the deep unfolding algorithm in cell-free massive MIMO systems. The simulation setup includes $B=4$ BSs, each with $M=4$ antennas. They serve $K=8$ UEs, randomly divided into $G=4$ multicast groups, with 2 UEs per group. The channel follows an uncorrelated Rayleigh fading model. Traditional precoding optimization methods often face challenges of high computational complexity and limited scalability in large-scale systems. In contrast, the deep unfolding algorithm extends conventional fixed-iteration algorithms into a hierarchical structure similar to neural networks, introducing learnable parameters at each layer to accelerate convergence and improve performance\cite{4}. To address this issue, we introduce a WAI optimization framework based on deep unfolding. The projected gradient descent (PGD) method is unfolded into a hierarchical structure, with trainable parameters introduced at each layer. This design enables rapid convergence within a limited number of iterations.
Unlike black-box DNNs, which rely solely on end-to-end data mapping, our approach explicitly inherits the mathematical derivation of the optimization algorithm. Each optimization step is formulated with a well-defined theoretical interpretation\cite{4}. 

Simulation results demonstrate that the deep unfolding PGD method, by incorporating adaptive step sizes, exhibits significant advantages over fixed-step methods. This method improves iteration efficiency, reduces the number of iterations required for convergence, and enhances overall system performance. Furthermore, the deep unfolding PGD method maintains strong robustness under imperfect CSI conditions, ensuring the stability of the system in practical applications.

\section{Conclusions and Future Directions}\label{sec5}
This article provides an overview of the fundamentals, advantages and applications for WAI-aided wireless communications. We have first introduced the basic theory and principles of WAI models. Then, we have explored the application scenarios in wireless communication. Finally, we demonstrated the robustness and superior performance of WAI models through case studies. Some other directions for future research in WAI-aided wireless communication systems are outlined as follows.


\textbf{Privacy and Security:} 
The theory-driven WAI model enhances the interpretability of the optimization process but also increases the risk of privacy leakage and security threats. Future research should explore differential privacy mechanisms to reduce the model’s reliance on individual data points, thereby mitigating model inversion attacks and membership inference attacks. Additionally, adaptive privacy mechanisms should dynamically regulate information-sharing strategies based on access privileges and data sensitivity. This approach aims to improve the traceability of the optimization process while reducing the risk of sensitive information leakage. Achieving a balance between privacy protection, interpretability, and communication system optimization will be essential for promoting the secure and reliable deployment of WAI in sensitive applications.

\textbf{Mobile Edge Intelligence:} 
The future trajectory of WAI and communication edge intelligence holds transformative potential for the deployment and reliability of large language models (LLMs) within mobile edge networks. WAI is poised to address delay-sensitive problems by advancing methodologies that elucidate the intricate operations of LLMs, enabling edge devices and servers to deliver outputs that are both comprehensible and trustworthy to end-users and regulatory bodies. Concurrently, communication edge intelligence will evolve to optimize the balance between computational efficiency and communication overhead, harnessing 6G network advancements to support real-time model updates and inference at the network edge.

\bibliographystyle{IEEEtran}

\bibliography{IEEEabrv,ref}
\end{document}